\documentclass[aps,pra,twocolumn,showpacs,letterpaper,superscriptaddress]{revtex4-1}

\usepackage[colorlinks=true, citecolor=blue, linkcolor=blue, urlcolor=blue]{hyperref}

\usepackage{graphicx,dcolumn,longtable,epsfig}
\usepackage[usenames]{color}
\usepackage{amssymb}
\usepackage{amsmath}
\usepackage{epstopdf}
\usepackage{bm}
\usepackage{footnote}
\usepackage{float}
\usepackage{subfigure}
\usepackage{color}
\usepackage{ulem}

\input epsf.tex

\def\bea{\begin{eqnarray}}
\def\eea{\end{eqnarray}}
\def\be{\begin{equation}}
\def\ee{\end{equation}}

\begin{document}
\title{Phase diagrams of the disordered Bose-Hubbard model with cavity-mediated long-range and nearest-neighbor interactions}
\author{ Chao Zhang}
\affiliation{Theoretical Physics, Saarland University, 66123, Saarbr$\ddot{u}$cken, Germany}
\author{Heiko Rieger}
\affiliation{Theoretical Physics, Saarland University, 66123, Saarbr$\ddot{u}$cken, Germany}

\begin{abstract}
Recent experiments with ultracold atoms in an optical lattice have realized cavity-mediated long-range interaction and observed the emergence of a supersolid phase and a density wave phase in addition to Mott insulator and superfluid phases. Here we consider theoretically the effect of uncorrelated disorder on the phase diagram of this system and study the two-dimensional Bose-Hubbard model with cavity-mediated long-range interactions and uncorrelated diagonal disorder. We also study the phase diagram of the extended Bose-Hubbard model with nearest-neighbor interactions in the presence of uncorrelated diagonal disorder. The extended Bose-Hubbard model with nearest-neighbor interactions has been realized in the experiment using dipolar interaction recently. With the help of quantum Monte Carlo simulations using the worm algorithm, we determine the phase diagram of those two models. We compare the phase diagrams of cavity-mediated long-range interactions with nearest-neighbor interactions. We show that two kinds of Bose glass phases exist: one with and one without density wave order. We also find that weak disorder enhances the supersolid phase.
\end{abstract}

\pacs{}
\maketitle

\section{Introduction}
\label{sec:sec1}

The interplay between disorder and interaction attracts a lot of attention in condensed matter and statistical physics. A certain degree of disorder is ubiquitous in all condensed matter, but a thorough understanding of these systems is impeded by a poor control over the disorder and competing interactions. On the other hand, ultracold atoms, especially bosons in an optical lattice become an important way to simulate condensed matter systems~\cite{Inguscio2010, DeMarco2009PRL, DeMarco2010Nature, Schneble2011, DErrico:2014gc, Rapsch:1999bw, Ceperley1991, Soyler:2011ik, Ceperley2011, Zhang:2015it, Zhang:2017ei, Niederle:2013jy, Zhang:2018ew}. In these experiments, interactions and disorder can be tuned independently. The short-range interaction can be realized using Feshbach resonances, while the long-range interactions have been studied using ultracold gases of particles with large magnetic or electronic dipole moments~\cite{Booth:2015fp, DePaz:2013ff, Lu:2011hl},  polar molecules~\cite{Yan:2013fn, Hazzard:2014bx}, atoms in Rydberg states~\cite{Saffman:2010ky, Lw:2012ct, Gunter:2013fv}, or cavity-mediated interactions~\cite{Baumann:2010js, Landig:2016il}. Random potentials are usually produced using speckle patterns~\cite{CreatDisorder1, DeMarco2009PRL, AAspect2012}, while quasi-periodic potentials can be generated using bichromatic lattices~\cite{Fallani:2007ir}. Other possibilities to engineer disorder include the introduction of localized atomic impurities~\cite{Schneble2011} and holographic techniques which produce point-like disorder~\cite{Morong:2015gha}. 

Theoretically, a paradigmatic model to describe the interacting bosonic particles in an optical lattice is the Bose-Hubbard model (BHM). The BHM without disorder and only on-site repulsion features two phases: a superfluid (SF) phase and a Mott insulator (MI) phase. The so-called extended BHM includes nearest-neighbor and/or long-range interactions, as for instance dipolar interactions and cavity-mediated interactions. The phase diagram of the extended BHM with dipolar interactions has been calculated in Ref~\cite{CapogrossoSansone:2010em, Zhang:ws}. In contrast to dipolar interactions which decays as $1/r^3$, cavity-mediated long-range interactions are global, which means that the interaction strength between two bosons does not decay with the distance between them. The ground state phase diagram of the extended BHM with cavity-mediated long-range interactions has been investigated extensively with the help of mean-field theory~\cite{Keller_2017, Li:2013bn, Niederle:2016fi, Dogra:2016hy, Chen:2016kv}, Gutzwiller ansatz~\cite{Sundar:2016ie, Flottat:2017gn}, quantum Monte Carlo~\cite{Habibian:2013eh, Dogra:2016hy, Flottat:2017gn, Habibian:2013kw}, Variational Monte-Carlo~\cite{Bogner:2019ij}, and exact diagonalization~\cite{Blass:2018iw, Igloi:2018ig} in 1D, 2D, and 3D. The results show that by adding cavity-mediated long-range interactions, the extended BHM exhibits a richer phase diagram with additional density wave (DW) and supersolid (SS) phases.

Introducing disorder into the standard BHM leads to the emergence of the gapless Bose glass (BG) phase, characterized by finite compressibility and absence of off-diagonal long-range order, always intervenes between the SF phase and MI phase~\cite{Pollet2009BG, Gurarie:2009it}. The phase diagram of the disordered extended BHM with nearest-neighbor interactions was calculated for 3D~\cite{Lin:2017cz, Kemburi:2012jw}, and the phase diagram of the disordered BHM with dipolar interactions was calculated for 2D~\cite{Zhang:2017ei}. 
However, the study of the extended BHM with cavity-mediated long-range interactions in the presence of disorder is still lacking. Whether the disordered potential enhances or suppresses the DW and SS phases here is still unknown.

In this paper, we use quantum Monte Carlo simulations based on the worm algorithm~\cite{Prokofev:1998gz} to study the phase diagram of the two-dimensional Bose-Hubbard model with cavity-mediated long-range interactions and uncorrelated disorder. The paper is organized as follows: in section~\ref{sec:sec2}, we introduce the Hamiltonian of the system with cavity-mediated long-range and nearest-neighbor interactions. In section~\ref{sec:sec3}, we discuss various phases and the corresponding order parameters. In section~\ref{sec4.1}, we present the phase diagrams of the 2D extended BHM with cavity-mediated long-range interactions and uncorrelated disorder. On the mean-field level the BHM with cavity-mediated interactions is identical to the BHM with nearest-neighbor repulsion - up to a renormalization of the chemical potential and the on-site potential~\cite{Dogra:2016hy}. Therefore, for comparison, we study in section~\ref{sec4.2} the phase diagram of the extended BHM with nearest-neighbor interactions in the presence of uncorrelated disorder. The extended BHM with nearest-neighbor interactions was experimentally realized in~\cite{Baier:2016ga}. Finally, section~\ref{sec:sec5} concludes this paper. 





\section{Hamiltonian} 
\label{sec:sec2}
In the following, we consider bosons trapped in an optical lattice with both short-range on-site and cavity-mediated long-range interactions in the presence of disordered potential. 
The system is described by the Hamiltonian~\cite{Habibian:2013eh, Habibian:2013kw, Niederle:2016fi}:
\begin{align}
\nonumber H& =-t\sum_{\langle i, j\rangle }(a_i^\dagger a_j+a_i a_j^{\dagger})+\frac{U_s}{2}\sum_i n_i(n_i-1) \\
&-\frac{U_l}{L^2} \Big{(}\sum_{i \in e} n_i - \sum_{j \in o} n_j \Big{)} ^2+ \sum_i (\varepsilon_{i}-\mu)n_i \;\; ,
\label{Eq1}
\end{align}
where the first term is the kinetic energy characterized by the hopping amplitude $t$. Here $\langle \cdots \rangle$ denotes nearest neighboring sites on an underlying square lattice of linear size $L$ with periodic boundary conditions, $a_i^\dagger$ ($a_i$) are bosonic creation (annihilation) operators satisfying the bosonic commutation relations. The second term is the short-range on-site repulsive interaction with interaction strength $U_s$. Here, $n_i=a_i^{\dagger}a_i$ is the particle number operator. The third term is the cavity-mediated long-range interaction with interaction strength $U_l$, the summations $i \in e$ and $j \in o$ denote summing over even and odd lattice sites respectively~\cite{Niederle:2016fi}. The fourth term is the chemical potential term with chemical potential $\mu$ shifted by the on-site random disordered potential $\varepsilon_{i}$, where $\varepsilon_{i}$ is uniformly distributed within the range $[-\Delta, \Delta]$. $\Delta$ is the disorder strength. We set the unit of energy and length to be the hopping amplitude $t$. For each  $U_l/t$, $U_{s}/t$, and $\Delta/t$, we average over 100-200 realizations of disorder.

On the mean-field level the BHM with cavity-mediated interactions is identical to the BHM with nearest-neighbor repulsion - up to a renormalization of the chemical potential and the on-site potential~\cite{Dogra:2016hy}. In order to check this, we also consider the disordered 2D BHM with nearest-neighbor interactions defined by the Hamiltonian:
\begin{align}
\nonumber H& =-t\sum_{\langle i, j\rangle }(a_i^\dagger a_j+a_i a_j^{\dagger})+\frac{U_s}{2}\sum_i n_i(n_i-1) \\
&+U_{nn} \sum_{\langle i, j\rangle} n_i n_j+ \sum_i (\varepsilon_{i}-\mu)n_i \;\; .
\label{Eq2}
\end{align}
Here, the first term is the kinetic energy with hopping amplitude $t$. The second term is the short-range on-site interaction with the interaction strength $U_s$. The third term is the repulsive interaction with interaction strength $U_{nn}$ between bosons on nearest neighboring sites. The fourth term is the disordered potential term coupled with the chemical potential term. For each  $U_{nn}/t$, $U_s/t$, and $\Delta/t$, we average over 500-1000 realizations of disorder. 



\section{Phases and order parameters}
\label{sec:sec3}

In this section, we list the phases we find in model~\ref{Eq1} and~\ref{Eq2} and the corresponding order parameters in table~\ref{Table1}. Each of the phases listed in table~\ref{Table1} corresponding to a unique combination of the order parameters. Here, three order parameters are needed to separate those quantum phases: superfluid stiffness $\rho$, structure factor $S(\pi, \pi)$, and compressibility $\kappa$. 


\begin{table}[h]

\includegraphics[trim=0.5cm 6cm 7cm 3cm, clip=true, width=0.48\textwidth]{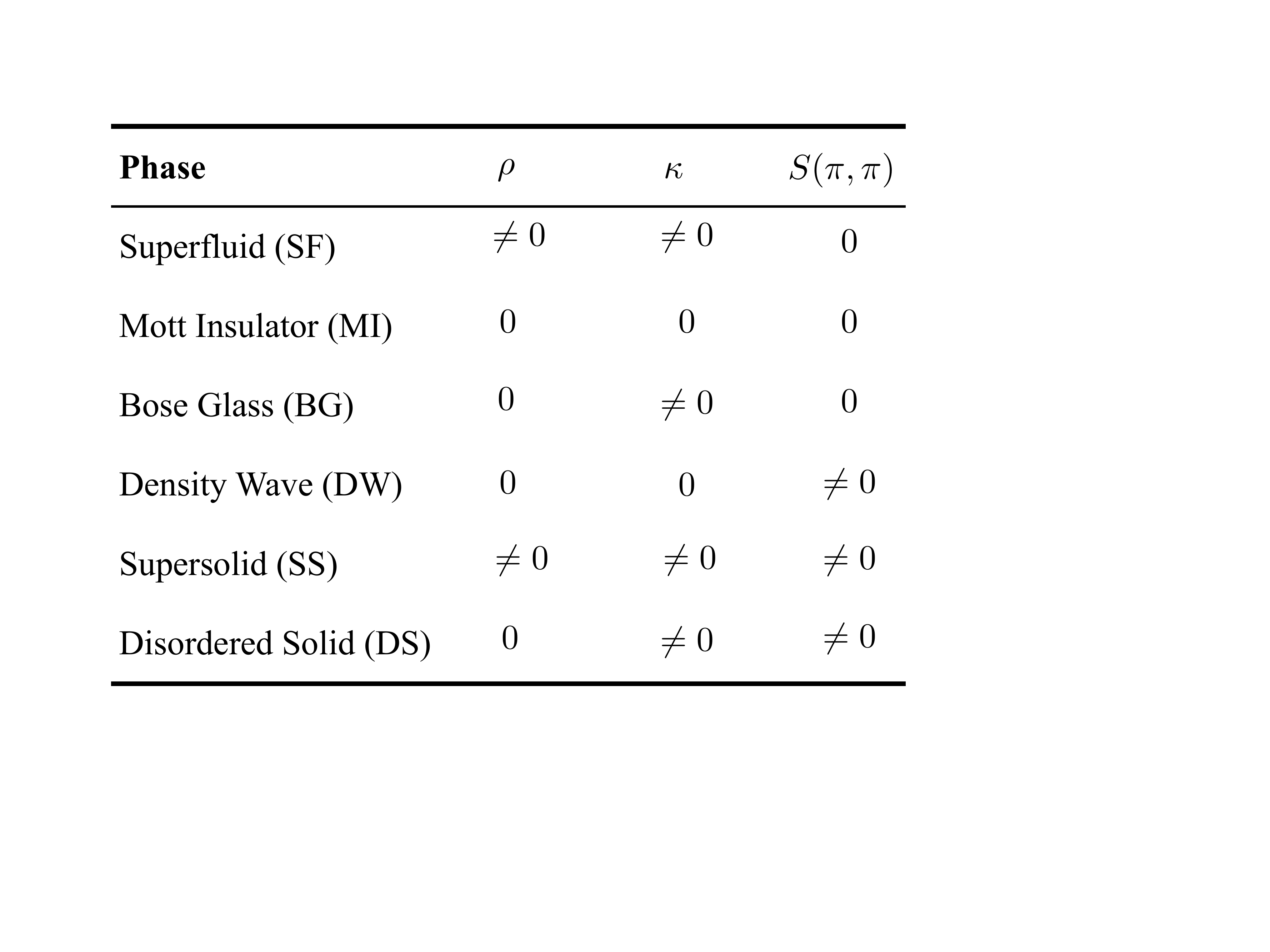}
\caption{Quantum phases and the corresponding parameters: superfluid stiffness $\rho$, structure factor $S(\pi, \pi)$, and compressibility $\kappa$.
}
\label{Table1}
\end{table}


A non-vanishing superfluid stiffness $\rho$ signifies off-diagonal long-range order, and it is easily accessible in QMC simulations using world line algorithms by calculation the winding number~\cite{Winding}. The superfluid stiffness is then given by:

\begin{equation}
\rho=\langle \mathbf{W}^2 \rangle /dL^{d-2}\beta \;\; .
\end{equation}
Here, $\mathbf{W}$ is the winding number. $d$ is the dimension of the system and here, $d=2$. $L$ is the linear system size and $\beta$ is the inverse temperature.
 
The structure factor characterizes diagonal long-range order and is defined as:
\begin{equation}
S(\mathbf{k})=\sum_{\mathbf{r},\mathbf{r'}} \exp{[i \mathbf{k} (\mathbf{r}-\mathbf{r'})]\langle n_{\mathbf{r}}n_{\mathbf{r'}}\rangle}/N \;\;.
\end {equation}
Here, $\mathbf{k}$ is the reciprocal lattice vector with $\mathbf{k}=(\pi, \pi)$ for the density wave with a checker board pattern and $N=L\times L$ is the system size.

The compressibility measures the density fluctuations and it is defined as:
\begin{equation}
\kappa=\beta (\langle n^2\rangle -\langle n \rangle ^2 ) \;\;.
\end{equation}

\section{Ground state phase diagrams}
\label{sec4}

\begin{figure*}[th]
\includegraphics[trim=0cm 0cm 0cm 0cm, clip=true, width=0.85\textwidth]{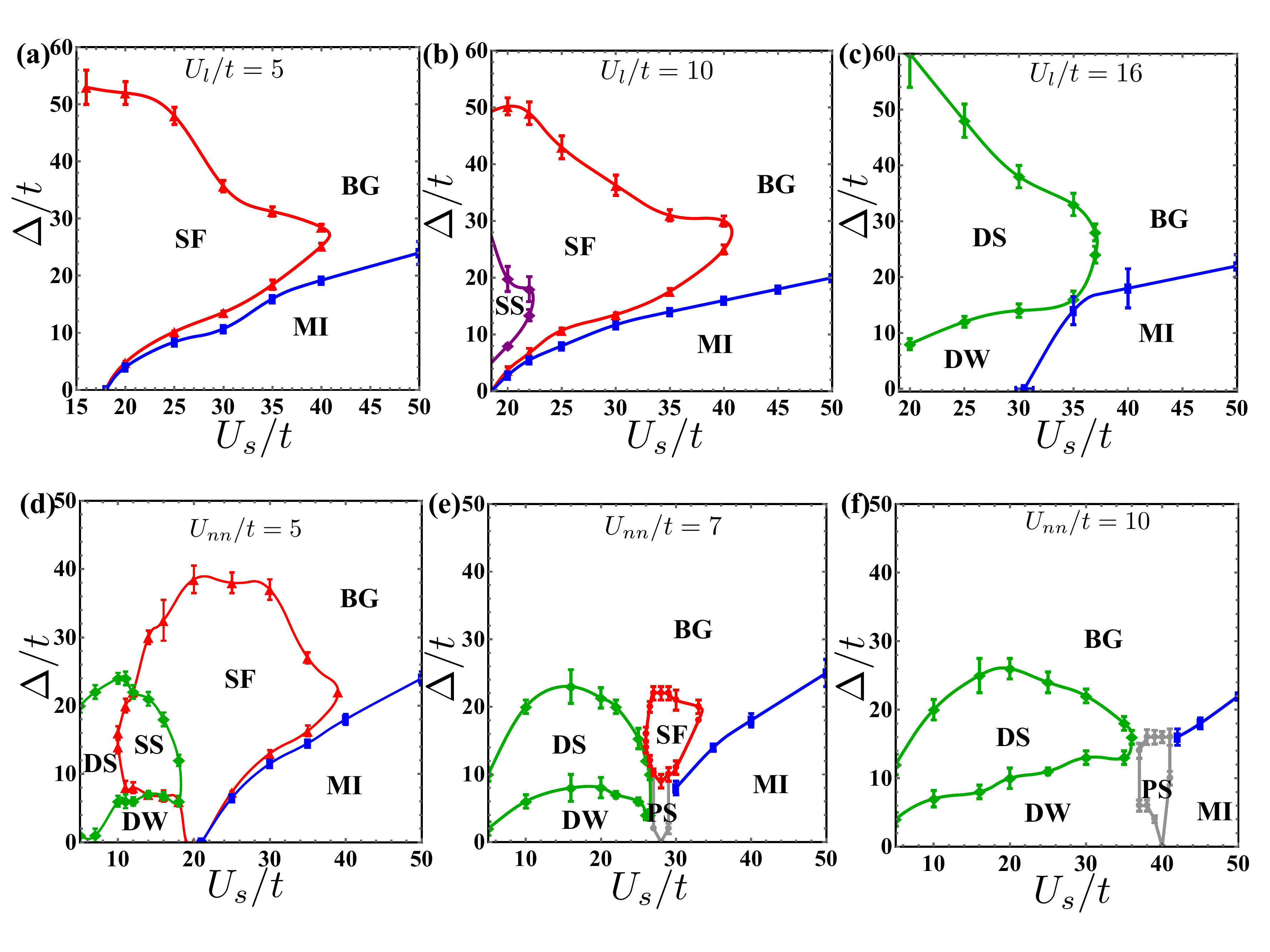}
\caption{(a)-(c) Ground state phase diagrams of model~\ref{Eq1} as a function of on-site interaction strength $U_s/t$ and disorder strength $\Delta/t$ at cavity-mediated long-range interaction $U_l/t=5$, $U_l/t=10$, and $U_l/t=16$, respectively. (d)-(f) Ground state phase diagrams of model~\ref{Eq2} as a function of on-site interaction strength $U_s/t$ and disorder strength $\Delta/t$ at nearest-neighbor interaction $U_{nn}/t=5$, $U_{nn}/t=7$, and $U_{nn}/t=10$, respectively. Here, PS indicates a region of phase separation.} 
\label{FIG1}
\end{figure*}

In this section, we present the ground state phase diagram for fixed particle density $\langle n_i \rangle=1$ (note that in this case the chemical potential in model (1) and (2) is superfluous) for cavity-mediated long-range interactions (Fig.~\ref{FIG1} (a)-(c)) and nearest-neighbor interactions (Fig.~\ref{FIG1} (e)-(g)). The x-axis is the on-site interaction $U_s/t$ and the y-axis is the disorder strength $\Delta/t$, here we set the hopping amplitude $t=1$. Table~\ref{Table1} summarizes the quantum phases in Fig.~\ref{FIG1} and the corresponding order parameters: superfluid stiffness $\rho$, structure factor $S(\pi, \pi)$, and compressibility $\kappa$. 
The phase boundary is determined by considering cuts through the x-axis ($U_s/t$) and calculating the above three order parameters as a function of disorder strength $\Delta/t$, as illustrated in Fig.~\ref{FIG2} and~\ref{FIG4}.

\subsection{Long-range interaction}
\label{sec4.1}

\begin{figure}[h]
\includegraphics[trim=3cm 4.4cm 18cm 0cm, clip=true, width=0.35\textwidth]{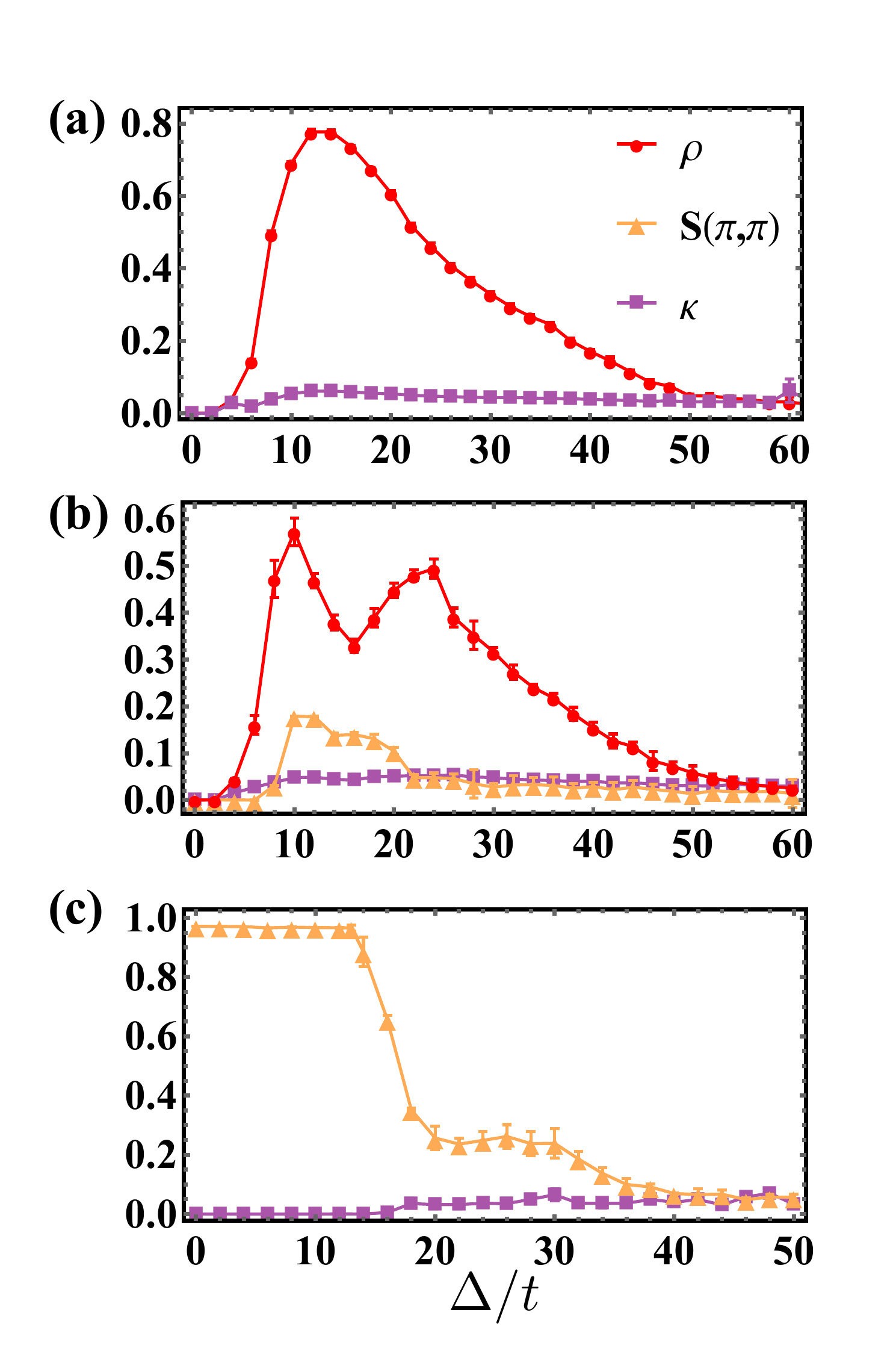}
\caption{Model 1 (cavity-mediated long-range interactions): superfluid stiffness $\rho$ (red circles), structure factor $S(\pi,\pi)$  (orange triangles), and compressibility $\kappa$  (purple rectangles) as a function of disorder strength $\Delta/t$ at $U_s/t=20$ for $U_l/t=5$ (a), $U_s/t=20$ for $U_l/t=10$ (b), and $U_s/t=30$ for $U_l/t=16$ (c). }
\label{FIG2}
\end{figure}

Figure~\ref{FIG1} (a)-(c) shows the phase diagrams of the disordered BHM with cavity-mediated long-range interactions at interaction strength $U_l/t=5$, 10, and 16 at filling factor $\langle n_i \rangle =1$. Without long-range interactions, the phase diagram of the disordered BHM at filling factor $\langle n_i \rangle=1$ contains three phases: an SF phase, a MI phase, and a BG phase~\cite{Soyler:2011ik}. We use system size $L=16$ and measure the three order parameters as a function of disorder strength $\Delta/t$ for various on-site interactions $U_s/t$ to determine the phase diagrams. Other system sizes have also been used to make sure the transition points are within the error bars. Figure~\ref{FIG2} (a) shows the superfluid stiffness $\rho$ and compressibility $\kappa$ as a function of disorder strength $\Delta/t$ at $U_s/t=20$ for $U_l/t=5$. The structure factor $S(\pi, \pi)$ is zero at fixed $U_s/t=20$ for different disorder strength. When the disorder strength $\Delta/t < 4$, the system is in the MI phase with zero superfluid stiffness and zero compressibility. At disorder strength $4<\Delta/t< 6$, the system is in the BG phase with finite compressibility but no superfluidity. As the disorder strength increases, the system goes to the SF phase at disorder strength $6<\Delta/t< 50$. Finally, at large disorder strength $\Delta/t>50$, the superfluidity is destroyed and the system enters the BG phase. Figure~\ref{FIG1} (a) shows the phase diagram at the interaction strength $U_l/t=5$. Compared with the phase diagram of the disordered BHM without long-range interactions, the shape of the phase boundaries at $U_l/t=5$ does not change but the region of the SF phase shrinks. For example, at $U_s/t=20$, the SF phase disappears around disorder strength $\Delta/t \sim 50$, while for the disordered BHM without long-range interactions, the SF phase exists up to $\Delta/t \sim 70$~\cite{Soyler:2011ik}. This is because the cavity-mediated long-range interaction tends to localize the particles in a `checkerboard' pattern which suppresses superfluidity.


Figure~\ref{FIG1} (b) shows the phase diagram at the interaction strength $U_l/t=10$. Compared with Fig.~\ref{FIG1} (a), the shape of the phase diagram boundaries does not change but with the SS phase emerges inside the SF phase at lower disorder strength. The SS phase has both diagonal long-range order and off-diagonal long-range order and is characterized by a finite superfluid stiffness $\rho$ and a finite structure factor $S(\pi, \pi)$. Figure~\ref{FIG2} (b) shows the three order parameters as a function of disorder strength $\Delta/t$ at $U_s/t=20$ for $U_l/t=10$. At disorder strength $9<\Delta/t < 20$, the system has a finite superfluid stiffness $\rho$ and a finite structure factor $S(\pi, \pi)$, implying that the system is in the SS phase. In the absence of disorder, at $U_l/t=10$, 
the DW to SF phase transition happens around $U_s/t \sim 15.5$~\cite{Bogner:2019ij}. 
Interestingly, by adding disorder to the system, the DW phase is transformed to the SS phase at weak disorder strength and the SS phase exists even around $U_s/t \sim 22$, implying that weak disorder enhances the SS order. Here, disorder transfers a solid into a percolating supersolid
~\cite{Lin:2017cz, Kemburi:2012jw} which is a percolating superfluid coexists with a solid.

\begin{figure}[h]
\includegraphics[trim=0.5cm 8cm 0cm 8cm, clip=true, width=0.5\textwidth]{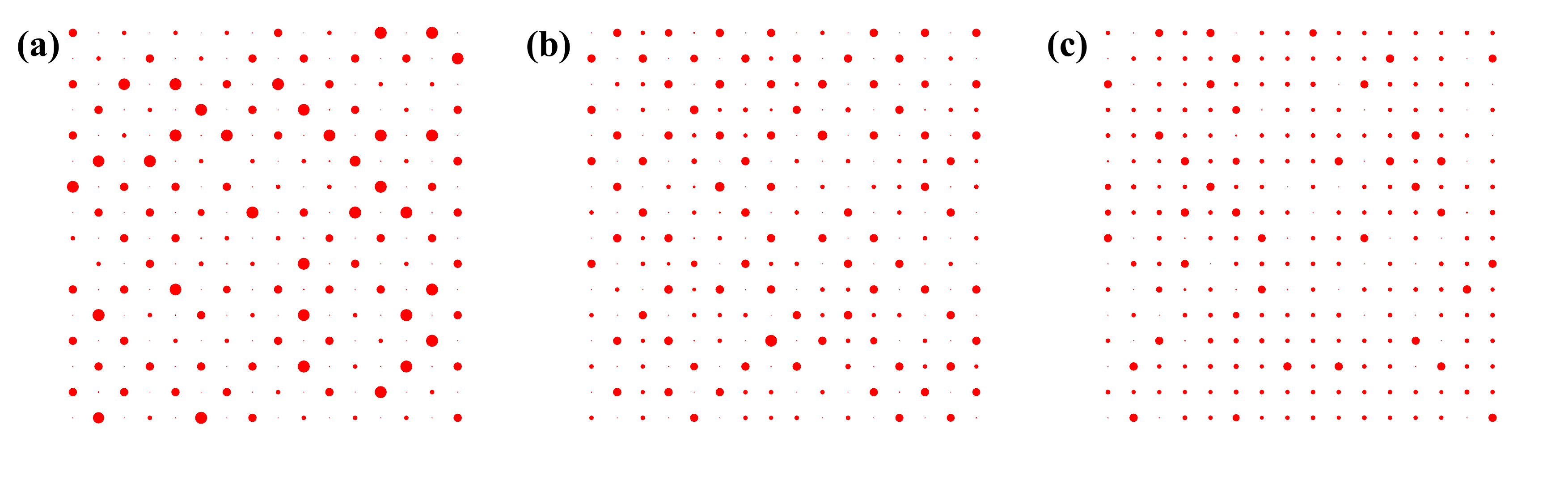}
\caption{Model 1 (cavity-mediated long-range interactions): density maps at $U_l/t=16$, $\Delta/t=26$, for different on-site interactions $U_s/t=25$ (a), $U_s/t=30$ (b), and $U_s/t=35$ (c), respectively. }
\label{FIG3}
\end{figure}

Figure~\ref{FIG1} (c) shows the phase diagram at the interaction strength $U_l/t=16$, where the superfluid phase has vanished and all bosons are localized. Interestingly, besides the DW and BG phase, a new glassy phase appears with finite compressibility and finite structure factor but no superfluidity. It is denoted as a disordered solid (DS). Figure~\ref{FIG2} (c) shows the three order parameters as a function of disorder strength $\Delta/t$ at $U_s/t=30$ for $U_l/t=16$. At disorder strength $16<\Delta/t < 32$, the system has finite compressibility $\kappa$ and finite structure factor $S(\pi, \pi)$, which shows that the system is in the DS phase. 
We find that, at lower $U_s/t$, we have the DW solid to DS phase transition first and then the DS to BG phase transition. This can be explained by the theory of inclusions~\cite{Pollet2009BG, Gurarie:2009it}, which states that a compressible glassy phase, the DS phase is surrounded incompressible phase, the DW phase. The glassy phase here is the DS since both of the DS phase and DW phase have finite structure factor. Figure~\ref{FIG3} shows the density maps at fixed $U_l/t=16$ and $\Delta/t=26$ for on-site interaction $U_s/t=25$, 30, and 35, respectively. The radius of a red circle at a given site is proportional to the density at that site. At on-site interaction $U_s/t=25$, the density map shows clearly the density wave pattern. As the on-site interaction increases, the system losses the density wave pattern and the structure factor decreases. By further increases the disorder strength, the DS phase is destroyed in favor of the BG phase.

For model~\ref{Eq1}, the region of parameter space corresponding to $U_s/t < 18$ has not been explored extensively. The reason is that the finite-size effects are much more pronounced at small on-site interactions, and at the same time, the long-range interaction term in the Hamiltonian slows down the worm update in the algorithm more and more when the system size increases such that systems sizes beyond L=16 are computationally inaccessible to us. 

\subsection{Nearest-neighbor interaction}
\label{sec4.2}

In this subsection, we study the ground state phase diagram of model~\ref{Eq2}. On the mean-field level the BHM with cavity-mediated interactions is identical to the BHM with nearest-neighbor repulsion - up to a renormalization of the chemical potential and the on-site potential - and the mean-field phase diagrams are identical~\cite{Dogra:2016hy}. Here we demonstrate that the true phase diagram, for 2D in the presence of disorder, are only vaguely similar and actually shows significant differences.

\begin{figure}[h]

\includegraphics[trim=3cm 1cm 20cm 0cm, clip=true, width=0.4\textwidth]{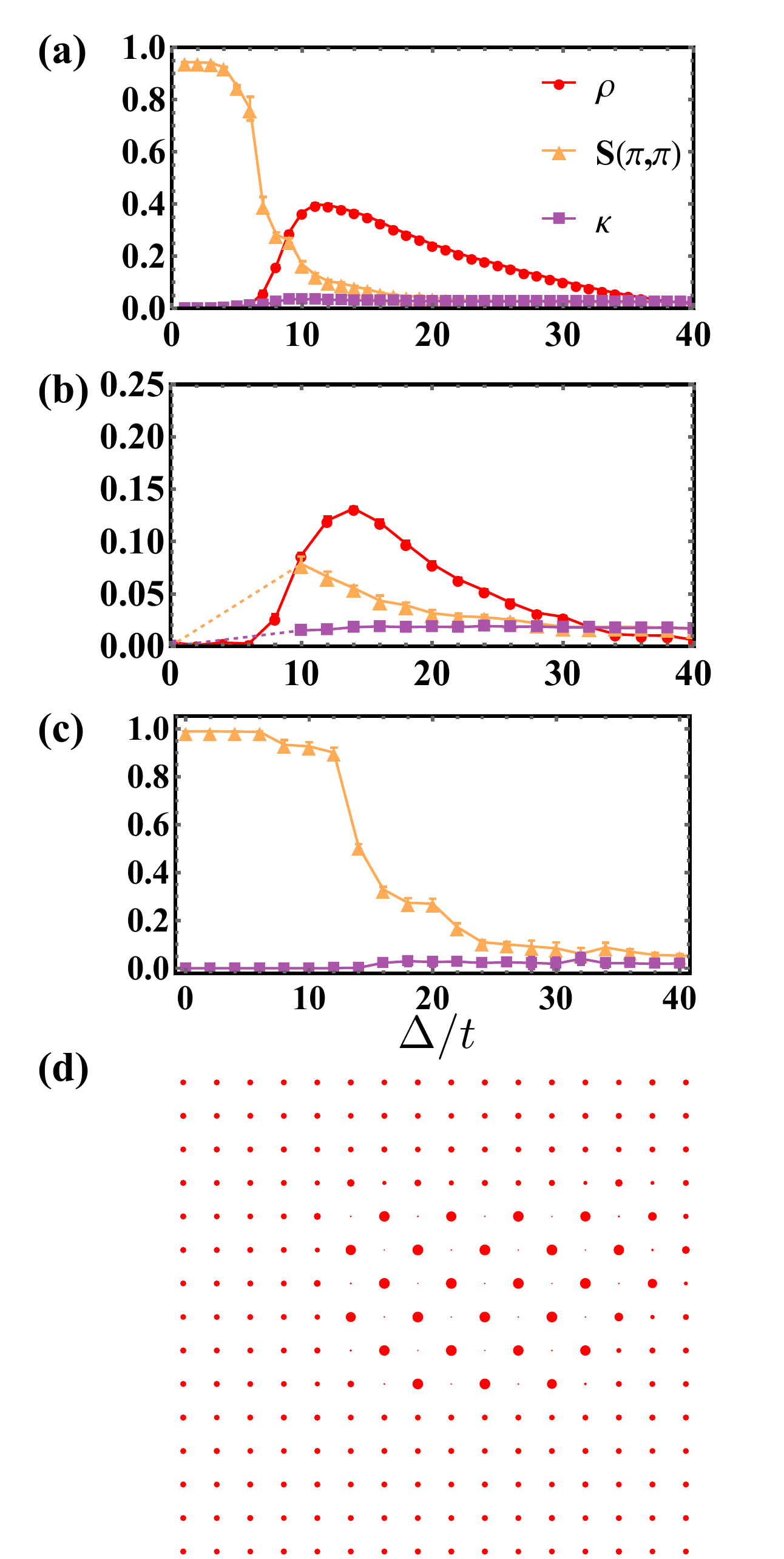}
\centering
\caption{Model 2 (nearest-neighbor repulsive interactions): superfluid stiffness $\rho$ (red circles), structure factor $S(\pi,\pi)$ (orange triangles), and compressibility $\kappa$ (purple rectangles) as a function of disorder strength $\Delta/t$ at $U_s/t=16$ for $U_{nn}/t=5$ (a), $U_s/t=28$ for $U_{nn}/t=7$ (b), and $U_s/t=30$ for $U_{nn}/t=10$ (c). The dotted line represents the PS region. (d) shows the density map at $U_s /t =28$ and $\Delta /t =6$.}
\label{FIG4}
\end{figure}

Figure~\ref{FIG1} (d)-(f) shows the phase diagrams of the disordered BHM with nearest-neighbor interaction at the interaction strength $U_{nn}/t=5$, 7, and 10 at filling factor $\langle n_i \rangle =1$, respectively. 
For the clean system without disorder, in the classical limit $U_s/t \rightarrow \infty$, the ground states are known~\cite{Sengupta:2005hq, BinXi:2011ig, Iskin:2011cr}. At filling factor $\langle n_i \rangle =1$, when the nearest-neighbor interaction and on-site interaction satisfy $zU_{nn}/U_s < 1$, the ground state is the MI state. While for $zU_{nn}/U_s > 1$, the ground state is the DW state. Here $z$ is the coordination number and $z=4$ in 2D.

Figure~\ref{FIG4} (a) shows the superfluid stiffness $\rho$, structure factor $S(\pi,\pi)$, and compressibility $\kappa$ as a function of disorder strength $\Delta/t$ at $U_s/t=16$ for $U_l/t=5$ and $L=16$. The structure factor decreases as the disorder strength $\Delta/t$ increases. At lower disorder strength $\Delta/t < 6$, the system is in the DW phase with zero superfluid stiffness, zero compressibility, but finite structure factor. At disorder strength $6<\Delta/t< 12$, the system is in the SS phase with finite superfluid stiffness and finite structure factor. Large disorder tends to destroy the DW order and the system goes in the SF phase at disorder strength $12<\Delta/t< 32$. Further increasing the disordered potential results in the destruction of the SF phase in favor of the BG phase. Figure~\ref{FIG1} (d) shows the phase diagram at interaction strength $U_{nn}/t=5$. When the on-site interaction $U_s/t >20$, there exists SF, MI, and BG phases, while when $U_s/t < 20$, there exists DW, SF, BG, and DS phases. For the clean system, the SS phase exists when the on-site interaction strength $U_s/t<5$~\cite{Bogner:2019ij}. At weak disorder strength, we find that the SS phase exists around $10< U_s/t <18$. Here, weak disordered potential enhances the SS phase~\cite{Lin:2017cz, Kemburi:2012jw}. 

Figure~\ref{FIG4} (b) shows the three order parameters as a function of disorder strength $\Delta/t$ at $U_s/t=28$ for $U_{nn}/t=7$. At lower disorder strength $0<\Delta/t<10$, the system is in a region displaying phase separation (PS). 
At larger disorder strength $\Delta/t>10$, the SF phase appears. The emergence of the SF phase for increasing disorder is due to the formation of a percolating SF cluster as described in~\cite{Kemburi:2012jw}.
Finally, strong disorder destroys the SF phase in favor of the BG phase. Figure~\ref{FIG4} (d) shows the density map of PS at $U_s /t =28$ and $\Delta /t =6$, where the DW phase is separated from the MI phase. 
Figure~\ref{FIG1} (e) shows the phase diagram at the interaction strength $U_{nn}/t=7$. Compared with the phase diagram in FIG.~\ref{FIG1} (d), as we increase the nearest-neighbor interaction, the SF phase shrinks and we find a region displaying PS at lower disorder strength around $U_s/t \sim 28$. 

Figure~\ref{FIG4} (c) shows the three order parameters as a function of disorder strength $\Delta/t$ at $U_s/t=30$ for $U_{nn}/t=10$. At disorder strength $16<\Delta/t < 24$, the system has finite compressibility $\kappa$ and finite structure factor $S(\pi, \pi)$, which shows that the system is in the disordered solid (DS) phase. Figure~\ref{FIG1} (f) shows the phase diagram at the interaction strength $U_{nn}/t=10$. At this interaction strength, there is no SF phase anymore and all bosons are localized. Phase separation occurs around $U_s/t \sim 40$ at lower disorder strength. Interestingly, in addition to the DW and BG phases, a disordered solid phase emerges. 
At lower $U_s/t$, the DW goes to DS and then BG phase as the disorder increases. The DS phase intervenes between the DW and BG phases since both the DW and DS phases have a finite structure factor.


\section{Conclusion}
\label{sec:sec5}

Comparing the phase diagrams of model~\ref{Eq1} and~\ref{Eq2}, we can see that the phase diagrams of the extended BHM with cavity-mediated long-range interactions and nearest-neighbor interactions only vaguely similar and actually display many significant differences. The main difference is that weak disorder leads to the PS in the nearest-neighbor interaction case. And in the weak disorder region, the phase diagram for the nearest-neighbor interaction changes around $zU_s/U_{nn}\sim 1$, where the DW dominates for $zU_s/U_{nn}<1$ while MI dominates for $zU_s/U_{nn} > 1$. There is no such change for the extended BHM with cavity-mediated long-range interaction. Here one always finds the DW phase for small $U_s$ and the MI phase for large $U_s$. In conclusion, the phase diagrams for the disordered BHM with cavity-mediated long-range and nearest-neighbor interactions are vaguely similar but with significant differences in the size of phases and the existence of the region of phase separation.


{\textit{Acknowledgements}}  
This work was performed with financial support from Saarland University. We would like to thank B. Capogrosso-Sansone for enlightening discussions. The computing for this project was performed at the OU Supercomputing Center for Education $\&$ Research (OSCER) at the University of Oklahoma (OU) and the cluster at Saarland University.

\bibliography{cavity}


\end{document}